\documentclass[fleqn,usenatbib]{mnras}

\usepackage{newtxmath} 
\usepackage{mathptmx}
\usepackage[T1]{fontenc} 
\usepackage{ae,aecompl}
\usepackage{txfonts}

\usepackage{amssymb}
\usepackage{graphicx}
\usepackage{amsmath}
\usepackage[dvips]{color}

\newcommand{\be}{\begin{equation}}
\newcommand{\ee}{\end{equation}}

\newcommand\eqnn[2]{
	\begin{equation}#2\label{#1}
	\end{equation}}

\title[Pulsars' braking indices and effective force]
  {Braking indices of pulsars obtained in the presence of an effective force}

\author[N. S. Magalhaes,  A. S. Okada and C. Frajuca]
  {N. S.~Magalhaes$^1$\thanks{Corresponding author.},
  A. S.~Okada$^2$ and C. Frajuca$^3$    \\
  $^1$Department of Exact and Earth Sciences, Federal University of Sao Paulo \\
	Rua Sao Nicolau 210, Diadema, SP 09913-030, Brazil\\
	$^2$ Scientific Initiation Program, Federal University of Sao Paulo \\
	Rua Sao Nicolau 210, Diadema, SP 09913-030, Brazil \\
	$^3$ Federal Institute of Education, Science and Technology of Sao Paulo\\
	R. Pedro Vicente 625, Sao Paulo, SP 01109-010, Brazil}

\date{Accepted XXX. Received YYY; in original form ZZZ}

\pubyear{2016}

\begin{document}

\label{firstpage}
\pagerange{\pageref{firstpage}--\pageref{lastpage}}
\maketitle

\begin{abstract}
Braking indices of pulsars present a scientific challenge as their theoretical calculation is still an open problem. In this paper we report results of a study regarding such calculation which adapts the canonical model (which admits that pulsars are rotating magnetic dipoles) basically by introducing a compensating component in the energy conservation equation of the system. This component would correspond to an effective force that varies with the first power of the tangential velocity of the pulsar's crust. We test the proposed model using data available and predict braking indices values for different stars. We comment on the high braking index recently measured of the pulsar J1640-4631.

\end{abstract}



\section{Introduction}

Pulsars are normally modelled as rapidly rotating, highly magnetized stars composed mainly of neutrons. It has been observed that their  rotation frequencies are decaying, this spin-down being quantified by the braking index (BI), $n$, defined by:
\eqnn{eq:def_n}{n  \equiv \frac{{\Omega \ddot \Omega }}
{{\dot \Omega ^2 }},}
where $\Omega$ is the pulsar's angular velocity and the dot denotes a time derivative. In such model, which we will refer to as  canonical, the main time-varying field responsible for the loss of rotational energy in a pulsar is a magnetic dipole field \citep{osg}. Also, the canonical model predicts n=3 for all existing pulsars.

There are not many pulsars for which the BI was obtained observationally (see Table \ref{tab:pulsars}). Most of their BI values lie within the range $0.9 - 2.8$ (see Table \ref{tab:bComplete}). The only pulsar with index greater than three is J1640-4631, whose value, $n=3.15$, was recently measured\citep{ar16}. Since the canonical model fails to yield the observed BI, improvements on this model have been tried involving different theoretical approaches \citep{br88, ah97, me97,  cs06, mag12, kt15}.


\begin{table*}
  \centering
  \caption{ Angular  velocities ($\Omega$) for pulsars with known braking indices. Time derivatives are denoted by a dot. }
	\label{tab:pulsars}
\begin{tabular}{cccccc}
\hline
PSR &  $\Omega$ &  $\dot{\Omega}$ ($\times 10^{-10}$ &$\ddot{\Omega}$ ($\times 10^{-21}$ & References \\
  &  (rad s$^{-1}$)&   rad s$^{-2}$) &  rad s$^{-3}$) & & \\
\hline
B0531+21 (Crab) & 189.912022 & -24.2674 &78.075 & \citet{lyne93,lyne15}  \\
B0833-45 (Vela) & 70.4 & -0.986 & 0.19  & \citet{lyne96,lyne15} \\
B0540-69 & 124.623817 & -11.8365 & 24.1  & \citet{liv2007} \\
B1509-58 &  41.68013054 & -4.24618765 & 12.2944  &\citet{liv2007}  \\
J1846-0258 & 19.340994108 & -4.21955 & 24.3  & \citet{liv2007} \\
J1119-6127 & 15.401361301 & -1.517708 & 4.014  & \citet{welte2011} \\
J1734-3333 & 5.37327178 & -0.104742 & 0.018 & \citet{espinoza2011} \\
J1833-1034 & 101.5322352 & -3.314451250 & 2.008734343 & \citet{rgl12}\\
J1640-4631 & 30.4320477075 & -1.433053 & 2.12 & \citet{ar16}\\
\hline
\end{tabular}
\end{table*}


In this paper we analyse a modification of the canonical model aiming at theoretically obtaining BI of pulsars that have already been observed. The modification consists basically on introducing a compensating component in the energy conservation equation of the system. This component would correspond to a force that varies with the first power of the tangential velocity of the pulsar's crust.

In the next section we present a summary of the canonical model followed by a description of its modified version as proposed by us. The remaining sections provide the results using the modified model and their analysis. We close the paper with our concluding remarks.


\section{Summary of the canonical model}

In the canonical model the energy carried by the radiation emitted by a pulsar results from magnetic energy (${E}_{mag}$), which, in turn, originates exclusively from the rotational kinetic energy ($E_{rot}$) of the neutron star, given by
\[
E_{rot} $=$ \frac{1}{2}I\Omega^2,
\]
where $I $=$ (2/5) MR^2$ is the moment of inertia of a solid sphere, assumed constant. The rotation power is thus
\begin{equation}
\dot{E}_{rot} $=$ I \; \Omega \; \dot{\Omega}.
\label{eq:RotPwr}
\end{equation}

For a rotating magnetic dipole, the radiated power is given by \citep{griff99, shap08},
\[
\dot{E}_{mag} $=$\frac{2}{3c^3}|\ddot{m}|^2,
\]
where $c$ is the speed of light in vacuum and  $\vec{m}$ is the dipole moment:
\[
\vec{m} $=$ \frac{B_PR^3}{2}$($\cos \alpha \hat{k}$+$\sin \alpha \cos $($\Omega t$)$ \hat{i}$+$\sin\alpha \sin$($\Omega t$)$ \hat{j}$)$.
\]
In the canonical model the following are constant: $B_P$ is the magnetic field at the pole, $R$ is the radius of the pulsar and $\alpha$ is the angle between the magnetic dipole axis and the rotation axis. The angular velocity of the pulsar, $\Omega$, varies with time. Therefore, the equation for the time-averaged radiated power becomes
\begin{equation}
\dot{E}_{mag} $=$ \frac{1}{6c^3}B_P^2R^6\Omega^4 \sin ^2\alpha.
\label{eq:MagPwr}
\end{equation}

Energy conservation implies
\begin{equation}
\dot{E}_{rot} $=$ - \dot{E}_{mag}
\label{eq:PwrCons}
\end{equation} 
which, using (\ref{eq:RotPwr}) and (\ref{eq:MagPwr}), yields
\begin{equation}
\dot{\Omega} $=$ - K\Omega^3,
\label{eq:spindown}
\end{equation}
where
\begin{equation}
K \equiv   \frac{{2 m_{\bot} ^2 }} {{3c^3 I}},
\label{eq:Km}
\end{equation}
is a constant, with $m_{\bot} \equiv \frac{B_PR^3}{2} \sin{\alpha}$.
Therefore, the canonical model predicts a gradual slowdown of the star's rotation. Moreover, as pointed out before, using (\ref{eq:spindown}) in (\ref{eq:def_n}) one finds $n $=$ 3$ for all pulsars.
 

\section{The proposed model}

In our model we will focus on the eight pulsars with BI less than three. Later we will comment on how the pulsar J1640-4631 might fit in it. For those eight pulsars, the canonical model is modelling a rotational energy dissipation larger than it actually is. Therefore, in the overall power balance in (\ref{eq:PwrCons}) an additional component may be assumed, its origin still to be determined. In this investigation we add the following effective force to the system corresponding to such component:
\[
\vec{F} $=$ b \vec{v},
\]
where $b$ is a positive constant with dimensions of $M/T$ and $\vec{v}$ is the tangential velocity of the pulsar's surface (at the radius $r $=$ R$). This functional form was chosen due to its simplicity as the force is only proportional to the first power of the velocity. We are thus assuming that the canonical model is basically correct except for a still unknown physical context mathematically modelled by this effective force. 

The work done by this force on the pulsar contributes to the energy balance and it is given by
\eqnn{eq:Fwork}{W $=$ \int^{l_1}_{l_2}\vec{F}.d\vec{l}.}

Since the force is parallel to the tangential displacement of the surface, $\vec{l}$, the dot product becomes an algebraic product. On the pulsar's surface $dl $=$ R \; d\phi$, where $\phi$ is the rotation angle around the rotation axis, whose time variation is the angular frequency: $\Omega $=$ d\phi /dt$. The tangential velocity of the pulsar's surface is in the direction of $\hat \phi$ and has modulus $v $=$ R\Omega$. Therefore, $\vec{F} $=$ b R \Omega \hat \phi$, implying 
\eqnn{eq:Fwork1}{W $=$ \int^{0}_{\phi} b R^2  \Omega d\phi \rightarrow W $=$ b R^2  \Omega \phi.}

The power associated with this work, $\dot{W}$ is then
\eqnn{eq:Fpwr}{\dot{W} $=$ b\, R^2 $($ \Omega^2 $+$ \dot \Omega \phi $)$.}

We now include this contribution in (\ref{eq:PwrCons}), obtaining 
\[
\dot{E}_{rot} $=$ - \dot{W}  -\dot{E}_{mag} 
\]
or
\begin{equation}
I \; \Omega \; \dot{\Omega} $=$ - b\, R^2 $($ \Omega^2 $+$ \dot \Omega \phi $)$  - K \; I \; \Omega ^4 .
\label{eq:PwrConsNew1}
\end{equation}

The expression for the BI in this model can be found differentiating (\ref{eq:PwrConsNew1}) with respect to time and then isolating $\ddot \Omega$, yielding
\begin{equation}
\ddot \Omega $=$ - $($ 3b \, R^2 \, \Omega \dot \Omega $+$ I \, \dot \Omega^2 $+$ 4K\, I \, \Omega^3 \dot \Omega$)$ $($ I \, \Omega $+$ b \, R^2  \phi $)$^{-1}.
\label{eq:ddotOmega}
\end{equation}

Substituting (\ref{eq:ddotOmega}) in (\ref{eq:def_n}) we found an expression for the BI that depends on $\phi$. This dependency can be eliminated with the aid of (\ref{eq:PwrConsNew1}) and we find the model's expression for the BI:
\begin{equation}
n $=$ 3 $+$ $($ K\, \Omega^3 $+$ \dot \Omega $)$ $($ K\, \Omega^3 $+$ \frac{b\,R^2 \Omega}{I} $)$ ^{-1}.
\label{eq:nNew}
\end{equation}
Note that when the force is absent, then $b$=0 and (\ref{eq:spindown}) is valid. Using these conditions in (\ref{eq:nNew}) results $n$=3, as it should be.


\section{Analysis of the tangential force}

The expression for the constant $b$ as a function of the BI is obtained from (\ref{eq:nNew}):
\begin{equation}
b $=$  \frac{K \, I \, \Omega^2}{R^2}  $+$ $($ - I \dot \Omega - K \, I\, \Omega^3 $)$ $($ $($ n-3 $)$ \Omega R^2 $)$^{-1}.
\label{eq:bComplete}
\end{equation}

We will estimate values for this constant based on observational data as well as on values for the other constants in the model. Observations indicate the existence of neutron stars with masses within a range \citep{demorest10}, but for the sake of estimates we adopted a typical value of \citep{lat04} $M $=$ 1.4 M_\odot$, where $ M_\odot $=$ 2 \times 10^{30} kg$ represents one solar mass.  Theoretical values for the star's radius vary from about 6 to 14 km \citep{lat04} so we adopted the usual value of $R $=$ 10^6$cm. We  chose $\sin ^2 \alpha $=$ 1$ to simplify the calculations. The speed of light in vacuum was approximated by $c $=$ 3.0 \times 10^{10}$ cm s$^{-1}$. As for the magnetic field, we adopted a typical value based on the canonical model: $B_P $=$ 10^{12}$G. 

We applied (\ref{eq:bComplete}) to the eight first pulsars in Table \ref{tab:pulsars} and obtained the results presented in Table \ref{tab:bComplete}, which show that $b$ is mainly of the order of 10$^{19}$ kg s$^{-1}$. With this estimate we looked back at the expression for the BI, (\ref{eq:nNew}), which we rewrite as
\begin{equation}
n $=$ 3 $+$ $($ 1 - \frac{ |\dot \Omega|}{ K\, \Omega^3 } $)$ $($ 1 $+$ \frac{b\,R^2 }{I K\, \Omega^2} $)$ ^{-1}.
\label{eq:nNew1}
\end{equation}




Assuming $b \sim 10^{19}$ kg s$^{-1}$, for all pulsars listed in Table \ref{tab:bComplete} we find $ \frac{ |\dot \Omega|}{ K\, \Omega^3 } >> 1$ and $\frac{b\,R^2 }{I K\, \Omega^2} >> 1$. This allows the approximation
\begin{equation}
n $=$ 3 - $($ \frac{ |\dot \Omega|}{ K\, \Omega^3 } $)$ $($ \frac{b\,R^2 }{I K\, \Omega^2} $)$ ^{-1} 
$=$  3 -   \frac{I }{R^2} \frac{ |\dot \Omega|/\Omega }{ b }  
$=$  3 -   \frac{2M }{5} \frac{ |\dot \Omega|/\Omega }{ b } .
\label{eq:nNew1_approx}
\end{equation}
This result shows a direct relation among $n$, $b$ and the ratio $|\dot{\Omega}| / \Omega$, which is different for the pulsars as shown in Table \ref{tab:bComplete}.


\begin{table*}
  \centering
  \caption{ The values for the force constant, $b$, were obtained from equation (\ref{eq:bComplete}) using observational values for the braking index, $n$, found in the references listed in Table \ref{tab:pulsars}. The values for the braking indices $n_{int}$ were obtained from the intersections of the fitted braking index function, (\ref{eq:nbFit}), and the braking index expression of the model, (\ref{eq:nNew1}). The respective value of the (n,b) pair at the intersecion is $b_{int}$. The percentual error ($\epsilon$) modulus compares $n_{int}$ to the observational value displayed in Table \ref{tab:bComplete}. The ratio $| \dot \Omega | / \Omega $ was calculated with data from Table \ref{tab:pulsars}. }
		\label{tab:bComplete}
\begin{tabular}{ccccccc}
\hline
PSR &  $b$ $\times 10^{19}$ & $n$ & $b_{int} \times$ 10$^{19}$ & $n_{int}$ & $| \epsilon |$ & $| \dot \Omega | / \Omega $  \\
   &  (kg s$^{-1}$) &  & (kg s$^{-1}$ ) & & (\%)&  $\times 10^{-12}$  (s$^{-1}$) \\
		\hline
Crab & 2.9678  & 2.51(1) & 2.3 (or 6.5) & 2.38 (or 2.78) & 5.1 (or 10.8) & 12.8 \\

Vela & 0.096714 & 1.4(2) & 0.084 & 1.10 & 21.4 & 1.4  \\
B0540-69 & 1.2378 & 2.140(9)  & 1.2 (or 8.2)  & 2.14 (or 2.87) & 0.0 (or 34.1) & 9.5 \\
B1509-58  &  7.22613 & 2.839(1)  & 7.8 (or 1.4) & 2.85 (or 2.19) & 0.6 (or 22.9) & 10.2  \\
J1846-0258 & 6.81415  & 2.65(1)  &  4.4 & 2.63 & 0.3 & 21.8\\
J1119-6127  &  3.49358 & 2.684(2)   & 7.9 (or 1.3) & 2.86 (or 2.16) & 6.4 (or 19.5) & 9.8 \\
J1734-3333 & 0.10201  & 0.9(2)  & 0.13 & 1.27 & 40.7 & 1.9\\
J1833-1034  & 0.31974    & 1.8569(6)  &  0.25  & 1.53  &  17.8 & 3.3 \\
\hline
\end{tabular}
\end{table*}


By inspecting the values of $n$ and $b$ in Table \ref{tab:bComplete} we noted that they are correlated despite the ratio $|\dot{\Omega}| / \Omega$: for stronger effective forces (higher $b$) the BI are higher. For this reason we fitted a curve with their values from that table and the function that we found, with R-squared coefficient of determination equal to 94.4\%, is:
\begin{equation}
n $=$ 0.3863849614 \, \ln $($ b $)$ - 14.8460540368,
\label{eq:nbFit}
\end{equation}
which implies
\begin{equation}
b $=$ 4.877253823 \times 10^{16} e^{2.588092446 \, n} .
\label{eq:bnFit}
\end{equation}

Since (\ref{eq:nbFit}) is optimized to be the best fit to the ($n$,$b$) pairs obtained from Table \ref{tab:bComplete}, its intersection (n$_{int}$ ,b$_{int}$) with (\ref{eq:nNew1})  is expected to be close to the actual value for the BI. We tested this hypothesis with the eight first pulsars of Table \ref{tab:pulsars} and (\ref{eq:nNew1}) can be intercepted in zero, one or two points. The results are presented in Table \ref{tab:bComplete}. When no intersection occurred (as for PSR J1846-0258, which has the higher value for the ratio $| \dot \Omega | /  \Omega $) we chose the (n$_{int}$ ,b$_{int}$) pair in the fitted function that corresponded to the shortest distance between the curves. 

When two intersections occurred and $|\dot \Omega | / \Omega$ was low (less than $ \sim$ $4\times$ 10$^{-12}$s), we chose as the best representative of the BI value, n$_{int}$, the one with lowest b$_{int}$-value; this was the case for Vela, PSR J1734-3333 and PSR J1833-1034. Otherwise we kept both values in Table \ref{tab:bComplete} knowing that the observational BI would be near one of the two the n$_{int}$ values; this was the case for the Crab pulsar, as well as the pulsars B0540-69, B1509-58 and J1119-6127.

The angular frequency can be correlated to the choice of one of these two values: when $\Omega$ is larger (smaller) than $\sim$ 100 Hz the lower (higher) value between the two n$_{int}$ values is closer to the observational one. For example, since the angular velocity of the Crab pulsar is high, then the lower value of n$_{int}$ is expected to be closer to the observational value (as it actually does). This choice is understandable when one inspects the values for $\ddot \Omega$ for these four pulsars; for instance, it is high for Crab, thus requiring a higher n$_{int}$ value for the corresponding $ |\dot \Omega |$/ $\Omega$.

\begin{figure}
\includegraphics[width=\columnwidth]{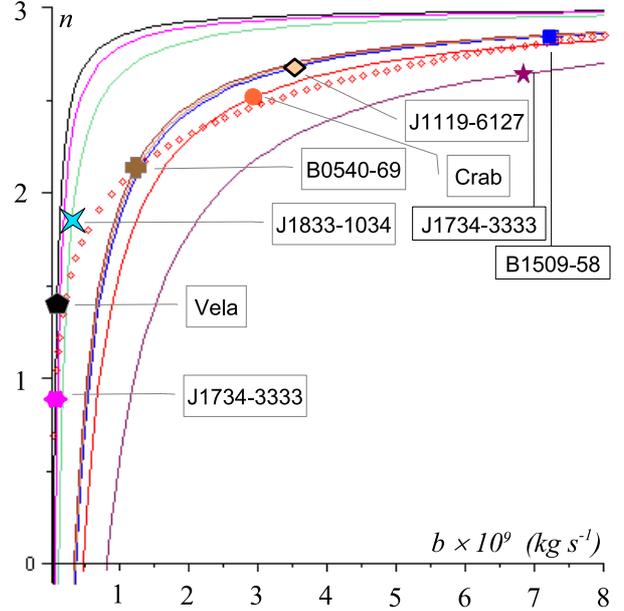}
\caption{ Plot of the dimensionless braking index, $n$, versus the effective force proportionality constant, $b$, for the pulsars listed in Table \ref{tab:bComplete}. The symbols are located at the values of $b$ obtained from the model trough equation (\ref{eq:bComplete}) using the observational values of $n$ presented in Table \ref{tab:bComplete}. That equation also provides the lines for each pulsar. The line made of small diamonds corresponds to the fitted function in equation (\ref{eq:nbFit}). The intersecion between this line and that of each individual pulsar is close to the respective symbol.}
\label{fig:Intersections}
\end{figure}

The intersection between the curves, shown in Figure \ref{fig:Intersections}, can be written algebraically through the substitution of (\ref{eq:bnFit}) in (\ref{eq:nNew1_approx}), yielding
\begin{equation}
 $($ 3 - n $)$ \exp $($ \frac{n}{0.3864} $) =$ 2.0565 \times 10^{-17} \frac{I}{R^2} \frac{ |\dot \Omega|}{\Omega }.
\label{eq:intersect}
\end{equation}

This equation, together with the procedures presented above for choosing n$_{int}$, allows us to predict BI for other pulsars. We show our predictions in Table \ref{tab:morePulsars}, whose pulsars have high value for $\dot \Omega$, thus being perhaps good candidates for the observational determination of $\ddot \Omega$. Also, those pulsars have ratios $ |\dot \Omega |$/$\Omega$ near the ones in Table \ref{tab:bComplete}. In Table \ref{tab:morePulsars} we include our BI prediction for the high-n pulsar J1640-4631, which we shall shortly comment on.


\begin{table*}
  \centering
  \caption{ Predicted braking indices for pulsars using our model. }
	\label{tab:morePulsars}
\begin{tabular}{cccccc}
\hline
PSR &  $\Omega$ &  $\dot{\Omega}$ ($\times 10^{-10}$ & $|\dot{\Omega}|$ / $\Omega$  & $n$  & References \\
  &  (rad s$^{-1}$)&   rad s$^{-2}$) & ($\times 10^{-12}$ s$^{-1}$) & (predicted) &  \\
\hline
J1418-6058 & 56.8 & -0.8702 & 1.53 & 1.14 & \citet{abdo09} \\
1E 1547.0-5408 & 3.03 & -34.0 & 11.2 & 2.83 & \citet{camilo07} \\
J1124-5916 & 46.4 & -2.576 & 5.56 & 2.94  & \citet{camilo02} \\
J1640-4631 & 30.4 & -1.433 & 4.71 & 2.95 & \citet{ar16} \\

\hline
\end{tabular}
\end{table*}


Our predicted BI values for J1418-6058 and 1E 1547.0-5408 (also known as J1550-5418) are in agreement with the predictions by \citet{mag12}. 

In Table \ref{tab:morePulsars} the numbers for pulsar J1124-5916 are close to those for J1640-4631, including our predicted BI. The value $n \sim 2.95$ is the largest that our model can provide. Since the latter pulsar has $ n =  3.15$, we wonder if the former would have $n > 3$ as well. As a generalization, perhaps pulsars that have BI close to 2.95 in our model might have $n>3$.This suggests that our model may be able to indicate pulsars with BI larger than three even though it was developed for pulsars with indices less than three.

\section{Concluding remarks}

We analysed a modification of the canonical model for pulsars' spin-down that introduces an effective force that is tangential to the star's rotation motion. Our goal was to provide more precise predictions of pulsars' BI. The effective force involves all yet unknown physical contributions that make pulsars' BI less than three. Our results were possible assuming that the pulsars have the same (typical) values for some physical characteristics, like mass and radius.

We used eight pulsars with observed BI to calibrate the model. Also, we discovered an extra relation between the BI, the ratio $|\dot \Omega|/\Omega$  and the tangential force constant of a pulsar which enabled us to make  predictions of BI of other pulsars.

The results that we found are applicable to  pulsars with $|\dot \Omega|$ larger than 10$^{-11}$ rad s$^{-2}$ and with $|\dot \Omega|/\Omega$ near the range 1 - 25 $\times$ 10$^{-12}$ s$^{-1}$. In order to improve the model, it is important to find physical details about the effective force. Its mathematical structure is simple and general, allowing different physical possibilities for its origin.

By using in our model data of the high-n pulsar J1640-4631 we found evidence that the model can also indicate pulsars with BI larger than three.

$ \\$
N.S.M., A.S.O. and C.F. acknowledge the Brazilian federal funding agency CNPq for financial support (grants 309295/2009-2, 149107/2010-2 and 312906/2013-7, respectively) as well as the  National Institute of Science and Technology in Astrophysics (INCT-A, Brazil) and FAPESP (thematic project, grant 13/26258-4).


\label{lastpage}


\begin{thebibliography}{}

\bibitem[\protect\citeauthoryear{Abdo et al.}{2009}]{abdo09}
Abdo A. A., Ackermann M., Ajello M. et al., 2009, Science, 325, 840

\bibitem[\protect\citeauthoryear{Allen \& Horvarth}{1997}]{ah97}
     Allen M. P., Horvath J. E., 1997, ApJ, 488, 409

\bibitem[\protect\citeauthoryear{Archibald et al.}{2016}]{ar16}
Archibald R. F. et al., 2016, ApJ, 819, L16

\bibitem[\protect\citeauthoryear{Blandford \& Romani}{1988}]{br88}
    Blandford R. D., Romani R. W., 1988, MNRAS, 234, 57P

\bibitem[\protect\citeauthoryear{Camilo et al.}{2002}]{camilo02}
Camilo F., Manchester R. N., Gaensler B. M., Lorimer D. R., Sarkissian
J., 2002, ApJ, 567, L71

\bibitem[\protect\citeauthoryear{Camilo et al.}{2007}]{camilo07}
Camilo F., Ransom S. M., Halpern J. P., Reynolds J., 2007, ApJ, 666, L93

\bibitem[\protect\citeauthoryear{Contopoulos \& Spitkovsky}{2006}]{cs06}
    Contopoulos I., Spitkovsky A., 2006, ApJ, 643, 1139

\bibitem[\protect\citeauthoryear{Demorest et al.}{2010}]{demorest10}
Demorest, P. B., Pennucci, T., Ransom, S. M., Roberts, M. S. E., Hessels,
J. W. T., 2010, Nature, 467, 1081

\bibitem[\protect\citeauthoryear{Espinoza et al.}{2011}]{espinoza2011}
Espinoza C.M., Lyne A. G., Kramer M., Manchester R. N., Kaspi V., 2011,
ApJ, 741, L13


\bibitem[\protect\citeauthoryear{Griffiths \& College}{1999}]{griff99}
Griffiths, D. J.; College, R., 1999, Introduction to electrodynamics, 3rd ed., Prentice Hall, Upper Saddle River


\bibitem[\protect\citeauthoryear{Kou \& Tong}{2015}]{kt15}
     Kou F. F., Tong H., 2015, MNRAS, 450, 1990	
		
\bibitem[\protect\citeauthoryear{Lattimer \& Prakash}{2004}]{lat04}
Lattimer, J. M., Prakash, M., 2004, Science, 304, 536
		
\bibitem[\protect\citeauthoryear{Livingstone et al.}{2007}]{liv2007}
   Livingstone M. A., Kaspi V. M., Gavriil F. P., Manchester R. N., Gotthelf E. V. G., Kuiper L., 2007, Astrophys. Space Sci., 308, 317

\bibitem[\protect\citeauthoryear{Lyne et al.}{1993}]{lyne93} 
  Lyne A. G., Pritchard R. S., Smith F. G., 1993, MNRAS, 265, 1003

\bibitem[\protect\citeauthoryear{Lyne et al.}{1996}]{lyne96}
   Lyne A. G., Pritchard R. S., Graham-Smith F., Camilo F., 1996, Nature, 381,
497

\bibitem[\protect\citeauthoryear{Lyne et al.}{2015}]{lyne15}
   Lyne A. G., Jordan, C., Graham-Smith F. et al., 2015, MNRAS, 446, 857

\bibitem[\protect\citeauthoryear{Magalhaes, Miranda \& Frajuca }{2012}]{mag12}
    Magalhaes N. S., Miranda T. A, Frajuca C., 2012, ApJ, 755, 54

\bibitem[\protect\citeauthoryear{Melatos}{1997}]{me97}
     Melatos A., 1997, MNRAS, 288, 1049

\bibitem[\protect\citeauthoryear{Ostriker \& Gunn}{1969}]{osg}
     Ostriker J. P., Gunn J. E., 1969, ApJ, 157, 1395
		
\bibitem[\protect\citeauthoryear{Roy, Gupta \& Lewandowski}{2012}]{rgl12} 
Roy J., Gupta Y., Lewandowski W., 2012, MNRAS, 424, 2213

\bibitem[\protect\citeauthoryear{Shapiro \& Teukolsky}{2008}]{shap08} 
Shapiro, S. L.; Teukolski, S. A., 2008, Black holes, white dwarfs and neutron stars:
the physics of compact objects, John Wiley $\&$ Sons, New York

\bibitem[\protect\citeauthoryear{Weltevrede, Johnston \& Espinoza}{2011}]{welte2011}
Weltevrede P., Johnston S., Espinoza C. M., 2011, MNRAS, 411, 1917


 
\end{thebibliography}
\end{document}